\documentstyle[12pt,a4,epsfig]{article}

\def\nextline{\hfill\break}
\def\nl{\nextline}
\def\mycomm#1{\nextline\strut\kern-3em{\tt ====> #1}\nextline}

\def\gray{\special{ps: 0.4 setgray}}
\def\black{\special{ps: 0.0 setgray}}

\newcommand{\draft}{
\newcount\timecount
\newcount\hours \newcount\minutes  \newcount\temp \newcount\pmhours
 
\hours = \time
\divide\hours by 60
\temp = \hours
\multiply\temp by 60
\minutes = \time
\advance\minutes by -\temp
\def\hour{\the\hours}
\def\minute{\ifnum\minutes<10 0\the\minutes
            \else\the\minutes\fi}
\def\clock{
\ifnum\hours=0 12:\minute\ AM
\else\ifnum\hours<12 \hour:\minute\ AM
      \else\ifnum\hours=12 12:\minute\ PM
            \else\ifnum\hours>12
                 \pmhours=\hours
                 \advance\pmhours by -12
                 \the\pmhours:\minute\ PM
                 \fi
            \fi
      \fi
\fi
}
\def\fullclock{\hour:\minute}
\begin{centering}
\gray
\special{ps: -90 rotate}
\special{ps: -4600 -5100 translate}
\font\Hugett  =cmtt12 scaled\magstep4
{\Hugett Draft: \today, \clock}
\black
\special{ps: 90 rotate}
\special{ps: 5100 -4600 translate}
\end{centering}
\vskip -1.7cm
$\phantom{a}$
} 

\newcommand{\bmath}{\begin{displaymath}}
\newcommand{\emath}{\end{displaymath}}
 
\def\beq{\begin{equation}}
\def\eeq{\end{equation}}
 
\newcommand{\bea}{\begin{eqnarray}}
\newcommand{\eea}{\end{eqnarray}}
 
\def\eqref#1{(\ref{#1})}
\def\bra#1{\left\langle #1\right|}
\def\ket#1{\left| #1\right\rangle}
 
\newcounter{saveeqn}
\newcommand{\alpheqn}{\setcounter{saveeqn}{\value{equation}}%
\stepcounter{saveeqn}\setcounter{equation}{0}%
\renewcommand{\theequation}
      {\mbox{\arabic{saveeqn}\alph{equation}}}}%
      
\catcode`\@=11 

\def\eqarraylabel#1{\@bsphack \if@filesw 
{\let \thepage \relax \def \protect {\noexpand \noexpand \noexpand }%
\edef \@tempa {\write \@auxout {\string \newlabel 
{#1}{{\mbox{\arabic{saveeqn}}}{\thepage }}}}\expandafter }\@tempa 
\if@nobreak \ifvmode \nobreak \fi \fi \fi \@esphack}

\catcode`\@=12 
 
\newcommand{\reseteqn}{\setcounter{equation}{\value{saveeqn}}%
\renewcommand{\theequation}{\arabic{equation}}}%
 
\def\tsize{\Large} 
\def\asize{\normalsize} 
\title {
\begin{flushright}
\normalsize TAUP-2572-99\\
\normalsize WIS-99/23/June-DPP\\
\end{flushright}
\vspace{2.0cm}
\tsize
Nucleon Spin with and without Hyperon Data:\\
\tsize
A New Tool for Analysis\thanks{Supported
in part by grant from US-Israel Bi-National Science Foundation}
}
\author{\asize
Marek Karliner$\,^{1}$~\footnote{\tt e-mail: marek@vm.tau.ac.il}\\
\asize and\\
\asize $\phantom{a}$ Harry J. Lipkin$\,^{1,2}$~\footnote{\tt e-mail:
ftlipkin@wiswic.weizmann.ac.il}
\vspace{0.5cm}
\\
\asize \sl $^1$\,School of Physics and Astronomy\\
\asize \sl The Raymond and Beverly Sackler Faculty of Exact Sciences\\
\asize \sl Tel Aviv University, 69978 Tel Aviv, Israel\\
\asize \sl and\\
\asize \sl $^2$\,Department of Particle Physics\\
\asize \sl The Weizmann Institute of Science, 76100 Rehovot, Israel\\
}
\date { }
\begin {document}
\maketitle
\begin{abstract}
We present a simple explanation of the underlying physics in the use
of hyperon decay data to obtain information about proton spin structure.
We also present an alternative input using nucleon magnetic moment data
and show that the results from the two approaches are nearly identical.
The role of symmetry breaking is clarified while pointing out that simple
models explaining the violation of the Gottfried sum rule via pion emission
tend to lose the good SU(3) predictions from Cabibbo theory for hyperon decays.
    \end{abstract}
\thispagestyle{empty} 
%
\newpage
\section{Introduction}

The conventional analyses of proton spin structure make use of three
experimental quantities to determine the values of the three contributions to
the proton spin from the three flavors of quarks denoted by $\Delta u $,
$\Delta d  $  and $\Delta s $. The connection between two of the commonly used
experimentally determined numbers to proton spin structure is reasonably clear
and well established. The use of the third, obtained from data on weak decays
of hyperons, rather than from data on the nucleon itself, involves assumptions
about SU(3) flavor symmetry relations between nucleon and hyperon wave
functions which have been challenged. 
 
In this note we wish to spell out this problem and indicate how SU(3) symmetry
breaking can be taken into account and also to present an alternative source for
the third experimental parameter, the ratio of the proton and neutron magnetic
moments, which depends only on the properties of the nucleon, and does
not require any assumptions about hyperon spin structure.
 
     Recent experiments of polarized deep inelastic scattering (DIS) provided us
 with
high quality data for the spin structure functions of the proton, deuteron and
 neutron~ \cite{SMC93a,E142,SMC94b,newpoldata}.
These measurements are used to evaluate the first moments of the spin dependent
structure functions which can be interpreted in terms of the contributions of
 the
quark spins ($\Delta \Sigma =\Delta u + \Delta d +\Delta s$) to the total spin
of the nucleon. The first results from the measurements of the proton spin
 structure function
by the EM Collaboration~ \cite{As88} were very surprising, implying that 
$\Delta \Sigma$ is
 rather small
(about 10\%) and that the strange $sea$ is strongly polarized. 
More recent analysis \cite{Erice95},\cite{mk99}, 
incorporating higher-order QCD corrections,
together with additional experimental data~\cite{SMC94b,newpoldata},
and new analyses of hyperon decays~\cite{Cl93,Hs88}, suggest that
 $\Delta \Sigma$
is significantly larger than what was inferred from the EMC experiment,
 concluding that
$\Delta \Sigma \approx 0.24 \pm 0.04$ and $\Delta s = -0.12 \pm 0.03$.

The polarized deep inelastic scattering experiments measure contributions 
$\Delta q$ of
individual quarks to the proton spin, weighted by the square of the
quark charge $e_q$,
\beq
\Gamma_1^p = \left(\,{1\over2} \sum_q e^2_q\,\Delta q \,\right)
\times \left[1 \,- \,
{\alpha_s\over\pi} \,+\,
{\cal O}\left(\left({\alpha_s\over\pi}\right)^2\right) \,+\,
\,\,\cdots\,\,\right] \qquad .
\label{Gamma1def}
\eeq
Thus they
measure a quantity proportional to $4 \Delta u + \Delta d + \Delta s$. 
The proportionality constant includes
perturbative QCD
corrections which are known to be significant also at higher orders.
The actual value of $\Gamma_1$ is subject to a considerable experimental
uncertainty, due to the unknown systematic error coming from the 
low-$x$ extrapolation beyond the measured region.
In principle, one also needs to keep in mind that
the higher-order corrections are different for the flavor
singlet and flavor non-singlet parts of $\Gamma_1$. In this paper we
focus on other issues, so for sake of simplicity we will
use the most recent E143 result \cite{newpoldata}
$\Gamma_1(Q^2=3\ \hbox{GeV}^2)
 = 0.132 \pm0.003(\hbox{stat.})\pm0.009(\hbox{syst.})$\,,
and only the leading-order QCD correction. However,
since all the perturbative QCD corrections
have the same sign as the leading-order, we shall use here an 
``effective" value
$\alpha_s=0.4$, which is higher than the standard value at
$Q^2=3\ \hbox{GeV}^2$. As we shall see, this simplification  yields results
which are still quite close to those of \cite{Erice95} and \cite{mk99}.
The
Bjorken sum rule tells us that the neutron weak decay constant
$g_A = \Delta u - \Delta d$. 

Between these two measurements we obtain 
the values of two linear combinations of $\Delta u (p)$, $\Delta d (p) $  and
$\Delta s (p)$,
\beq
4\Delta u + \Delta d + \Delta s = 2.72
\label{QQ1.1a}
\eeq
\beq
\Delta u - \Delta d = 1.26 \label{QQ1.1b}
\eeq
Since these are the experimental quantities obtained directly from the data on
deep inelastic scattering and $G_A/G_V$ without SU(3) symmetry assumptions, we
can use these as a base for our further analysis. We now need an additional
third experimental input to determine the values of $\Delta u $,
$\Delta d $  and $\Delta s $.
 
   Up to this point the only assumption made about the proton wave function is
that it has a good isospin and that the strangeness-conserving components of
the weak axial current are isovectors.
   In order to obtain a value for $\Delta s $ it is necessary to use 
additional data; e.g. the hyperon decay data commonly used.. However this 
requires additional assumptions about SU(3) flavor
symmetry which is known to be broken. We now examine
the underlying physics of this symmetry breaking.

Before we break the symmetry we need to know why we need it in the first place.
   We need isospin SU(2) symmetry in order to obtain information about proton
spin structure from the neutron decay
\beq
g_A\equiv {G_A \over G_V}(n \rightarrow p) = \Delta u(p) -\Delta d(p)=
\Delta d(n) - \Delta u(n)
\label{eII}
\eeq
where we have used the Bjorken sum rule which relates the deep-inelastic data
to $g_A$ and isospin to relate the charged and neutral strangeness-conserving
axial currents and to relate the proton and neutron wave functions. The neutron
and proton are isospin mirrors which go into one another under the
$u \leftrightarrow d$ transformation.
 
Similarly we use SU(3) symmetry if valid to obtain information about proton
spin structure from the $\Sigma^-$ semileptonic decay
\alpheqn 
\beq
{G_A \over G_V}(\Sigma^- \rightarrow n)  =  \Delta u(n) -\Delta s(n)=
\Delta d(p) -\Delta s(p)
\label{eIIIa}
\eeq
\beq
\phantom{{G_A \over G_V}}
\Delta s(\Sigma^-) -\Delta u(\Sigma^-) = 
\Delta u(n) - \Delta s(n) =
\Delta d(p) -\Delta s(p)
\phantom{{G_A \over G_V}}
\label{eIIIb}
\eeq
\beq
\phantom{{G_A \over G_V}}
|G_V(\Sigma^- \rightarrow n)|  = |G_V(n \rightarrow p)|
\phantom{{G_A \over G_V}}
\label{eIIIc}
\eeq
%
%
\eqarraylabel{eIII}
\reseteqn
where SU(3) relates the nucleon and $\Sigma^-$ wave functions. The neutron and
$\Sigma^-$ are SU(3) mirrors which go into one another under the
$u \leftrightarrow s$ transformation.
 
\noindent
Note that with precise data and SU(3) symmetry the value of
\hbox{$(G_A / G_V)(\Sigma^- \rightarrow n)$}
 is sufficient to give us all the
information needed for the spin structure of the proton. There is no need for
the F and D parametrization. It is only when we want to improve statistics by
also using other weak decays which involve the $\Lambda$ that we need F and D.
The $\Sigma^-(dds) \rightarrow n(ddu)$ is simple because the
strangeness-changing current at the quark level is an $s \rightarrow u$
transition which can only change the $\Sigma^-$ into a neutron. On the other
hand the same $s \rightarrow u$ transition on a $\Xi^-(dss)$ produces a
$(dsu)$ state with is a linear combination of a $\Lambda$ and a $\Sigma^o$.
How to separate this into the $\Lambda$ and $\Sigma^o$ requires an additional
parameter that depends on the hadron wave functions. It is conventional to use
the F and D parametrization for historical reasons, but there is no obvious
physical reason to use these parameters rather than any others.
For our purposes here it is sufficient to consider only the
$\Sigma^- \rightarrow n$ decay and see how the relations \eqref{eIII}
 are affected by
SU(3) symmetry breaking.
 
We see that there are four physical quantities that enter into this relation:
 
(1) $G_A(\Sigma^- \rightarrow n)   $
 
(2) $G_V(\Sigma^- \rightarrow n)  $
 
(3) $\Delta u(n) -\Delta s(n)$
 
(4) $\Delta s(\Sigma^-) -\Delta u(\Sigma^-) $.
 
When SU(3) is broken, these four quantities are no longer related, and we have
to understand what the breaking does to these relations. This depends upon how
SU(3) is broken. There is no model-independent way to allow for SU(3) breaking.
 
The quantity denoted by $g_A$ is really a
ratio of axial-vector and vector matrix elements. Both matrix elements
can be changed by SU(3) symmetry-breaking, but it is only the axial
matrix element that is relevant to the spin structure. The information
from hyperon decays used in conventional treatments of spin structure
is expressed in terms of D and F parameters which characterize the
axial couplings.  But there is the implicit assumption that the vector coupling
is pure F and normalized by the conserved vector current,
where the whole SU(3) octet of vector currents is conserved.
Thus any attempts to parameterize SU(3) breaking in fitting
hyperon data by defining ``effective" D and F parameters immediately
encounter the difficulty of how much of the breaking comes from the
axial couplings and how much comes from the vector
and the breakdown of the conserved vector currents for
strangeness changing currents. The vector matrix
element is uniquely determined by Cabibbo theory in the SU(3) symmetry
limit. The known agreement of experimental vector matrix elements
with Cabibbo theory places serious constraints on possible SU(3) breaking
in the baryon wave functions. On the other hand, the strange quark
contribution to the proton sea is already known from experiment
to be reduced roughly by a factor of two from that of a flavor-symmetric
sea \cite{CCFR}. This is expected to violate the $\Sigma^- \leftrightarrow n$
mirror symmetry since it is hardly likely that the strange sea should be
enhanced by a factor of two in the $\Sigma^-$. Yet Cabibbo theory requires
retaining the relation between the vector matrix elements \eqref{eIIIc}.

For insight into how to insert flavor asymmetry into procedures for obtaining
the spin structure of baryons from experimental data we first note that 
two mechanisms have been introduced for breaking flavor symmetry in the 
antiquark distributions in the nucleon. 

(1) Introducing a pion cloud, without other pseudoscalar mesons,
 while maintaining overall isospin 
symmetry \cite{piNfluct}.

(2) Reducing the strange contribution in the sea, thereby breaking SU(3)
    symmetry \cite{PJECH,HJL94}  .

In both cases, the question arises of whether this symmetry breaking is 
consistent with the experimental confirmation of the predictions from 
Cabibbo theory for the vector currents and the experimental agreement of the 
axial vector weak transitions with SU(3) symmetry relations.

The essential physics of the mechanism (1) is seen in the simple quark diagram 
for pion emission from a valence $u$ quark.     
\beq
u \rightarrow u + G \rightarrow u + \bar q q \rightarrow (u\bar q)_P + q
\label{WW1.1}
\eeq
where $G$ denotes gluons and  $(u\bar q)_P$ denotes a pseudoscalar meson with
the quark content $(u\bar q)$. Although flavor symmetry suggests that the
probabilities of producing $\bar u$ and $\bar d$ antiquarks via this diagram
must be equal, the constraint that the pseudoscalar meson constructed in this
way must be a pion leads to the result that the probability of 
producing a $\bar d$ antiquark is double that of producing a $\bar u$.
This factor of two can be seen by comparing the the $(u \bar d)_P$ and 
$(u\bar u)_P$ wavefunctions.
Whereas $(u \bar d)_P$ is a pure $\pi^+$,
the $(u\bar u)_P$ wave function is
a linear combination of the $\pi^o$, $\eta$ and $\eta'$ wave function with
a probability of only (1/2) of fragmenting into a $\pi^0$. Neglecting the 
$\eta$ and $\eta'$ contributions to a pion cloud model introduces a breaking
of nonet symmetry and SU(3) symmetry while conserving isospin. In this model
the neutron $\beta$ decay occurs both in the valence nucleon and the pion
cloud, and the isospin symmetry of the overall wave function preserves the 
conserved vector current and the Bjorken sum rule. The excess of $\bar d$ 
antiquarks over $\bar u$ can explain the observed violation of the Gottfried
sum rule \cite{Gottfried}. 

This mechanism breaks SU(3) and a simple toy-model calculation shows that it
can introduce serious disagreements with Cabibbo theory for 
strangeness changing transitions and in particular with the experimentally
verified predictions for hyperon decay.

To see this, we write the physical nucleon wave function as
a mixture of a ``bare" nucleon and a nucleon plus a pseudoscalar meson,
\beq
\ket{p_{phys}} = \cos (\phi) \cdot \ket{p_{val}} + 
{{\sin (\phi)}\over{\sqrt{3}}} 
\cdot \left[\,\ket {p\pi^o} - \sqrt{2}\cdot \ket {n\pi^+}\,\right] 
\label{Npi2}
\eeq
where $\ket {p_{val}}$ denotes the standard quark-model proton wave
function and $ \ket {N\pi}$ denotes a nucleon-pion wave function with angular
momentum $(J = 1/2)$ and isospin $(I = 1/2)$.
The factor $\sqrt 2$ which breaks the Gottfried sum rule appears here as an
isospin Clebsch-Gordan coefficient.

We now investigate the action of the strangeness-changing components of the
charged weak vector current on the proton wave function (\ref{Npi2}).
At zero-momentum transfer, these are just the $V$ spin raising and lowering 
operators, denoted by $V_\pm$, which generate
$u \leftrightarrow s$ and $\bar s \leftrightarrow \bar u$ transitions at
the quark level. The requirement that the proton and 
$\Lambda$ are members of the same SU(3) octet gives the two conditions:
\beq
V_+ \ket {p_{phys}} = 0
\label{V1}
\eeq
\beq
P(I=0) \cdot V_- \ket {p_{phys}} = {{\sqrt 6}\over{2}} \ket {\Lambda_{phys}}  
\label{V2}
\eeq
where $P(I=0)$ denotes a projection operator which projects out the $I=0$
component of the wave function and $\ket {\Lambda_{phys}} $ denotes the 
normalized physical $\Lambda$ wave function. These two conditions required by 
Cabibbo theory are manifestly violated by the proton wave function (\ref{Npi2})
except for the trivial case $\phi = 0$: (i) the left hand side of 
the condition (\ref{V1}) is a $pK^+$ state and does not vanish; 
(ii) the state 
$\ket {\Lambda_{phys}}$ defined by the condition (\ref{V2}) is not normalized
but satisfies
\bea
\bra {\Lambda_{phys}} \Lambda_{phys} \rangle &=& 
{{2}\over{\sqrt 6}} \cdot  \bra {\Lambda_{phys}} P(I=0) \cdot V_- 
\ket {p_{phys}} = \nonumber\\
\phantom{a}\label{V5}\\
 &=&\cos^2 (\phi) + (5/6) \sin^2(\phi) = 1 - (1/6) \sin^2(\phi) 
 \nonumber
\eea

The matrix element $\bra {\Lambda_{phys}} P(I=0) \cdot V_-\ket {p_{phys}}$ 
appearing in eq. (\ref{V5}) is just the transition matrix measured 
experimentally in the semileptonic vector $\Lambda \rightarrow p$ decay. 
Thus the inconsistency in eq.~(\ref{V5}) is not only a disagreement with 
Cabibbo theory; it is also a disagreement with experiment. 
The nature of this inconsistency is illuminated by noting that production of a
state of strangeness +1 by the action of the SU(3) generator $V_+$ 
when acting on a proton model wave function indicates that this proton
wave function is not a pure SU(3) octet but contains a {\bf 27} 
admixture. When SU(3) symmetry is restored in this model wave 
function by adding the correct admixture of  
$\Lambda \,K$, $\Sigma\, K$ and $p \,\eta_8 $ states in \eqref{Npi2}, the 
action of the operator $V_+$ on these components produces the $p\,K^+$ state 
with just the right phase to cancel the $p\,K^+$ state produced on the 
nucleon-pion state.

It is just these extra $\Lambda\, K$, $\Sigma\, K$ and $p \,\eta_8 $ components in
the nucleon wave function which are needed to restore the normalization of the
physical $\Lambda$ state $\ket {\Lambda_{phys}}$. This shows that if
baryon-meson components are added to the proton and $\Lambda$ wave functions,
the physical $\Lambda$ state is required by Cabibbo theory to decay also to
$\Lambda \,K$ and $\Sigma \,K$ components in the proton wave function. Leaving
these components out of the proton leads to disagreement with the semileptonic
$\Lambda \rightarrow p$ vector decay. If the $\Lambda$ wave 
function is evaluated explicitly from the lhs of eq. (\ref{V2}), it will include
$N\,K$ components with a kaon cloud, along with a
$\Sigma \pi$ component containing a pion cloud. If the kaon cloud is not
included, the disagreement in eq. (\ref{V5}) is much worse, with the
coefficient (1/6) replaced by (2/3). 

We thus see that in any model which includes a pion cloud in
the proton wave function, SU(3) breaking must reduce the kaon cloud
relative to the pion cloud from the value in the symmetry limit. This 
breaking seems to have a serious effect on the matrix elements of the 
strangeness changing current responsible for hyperon decays.
We will not address this issue further here.

A model which has been suggested \cite{PJECH} for breaking SU(3) via the 
mechanism (2) keeps all the good results of Cabibbo theory by introducing
a flavor asymmetric sea with no net flavor quantum numbers
into a baryon wave function whose 
valence quarks satisfy SU(6) symmetry and whose sea is {\em the same} for all 
baryons. 

The baryon wave function can be written,
\beq 
\Psi(B)= \psi_{val}(B) \cdot \phi_{sea}(Q=0) 
\label{WW1.2}
\eeq
where $\psi_{val}(B)$ denotes the valence quark wave function obtained from 
the SU(6) quark model and $\phi_{sea}(Q=0)$ denotes a sea with zero electric 
charge which may be flavor asymmetric.

The operation of any charged current operator $J_{\pm}$ on this baryon wave 
function is then    
\beq 
J_{\pm}\Psi(B)= \{J_{\pm}\psi_{val}(B)\} \cdot \phi_{sea}(Q=0) +
\psi_{val}(B) \cdot \phi'_{sea}(Q=\pm 1) 
\label{WW1.3}
\eeq
where $\phi'_{sea}(Q=\pm 1)  $ denote charged seas obtained by acting on the
neutral sea with the charged current. The exact structures of 
$ \phi'_{sea}(Q=\pm 1)  $ depend upon the details of the wave function, but are
irrelevant for our purposes. 
Since the overlaps of the identical neutral seas gives a factor unity and
the overlap of a neutral sea and a charge sea vanishes,
we see that the matrix elements of the 
charged current between any two baryon states $B$ and $B'$ is given by
\beq 
\bra{B'} J_{\pm} \ket{B} = \bra{B_{val}}  J_{\pm} \ket{B_{val}}
\label{WW1.4}
\eeq 
 We thus see that all
charged current matrix elements are given by the valence quarks. This provides
an explicit justification for the hand-waving argument \cite{PJECH} in the toy
model that in the hyperon decay the sea behaves as a spectator. In particular,
for the strangeness changing vector charge producing the $\Sigma^- \rightarrow
n$ decay, 
\beq 
\bra{n}  V_+ \ket{\Sigma^-}= 1               
\label{WW1.5}
\eeq
consistent with Cabibbo theory.

Unlike the charged current,
the matrix elements of the neutral components of the weak currents 
{\em do} have
sea contributions, and these contributions are observed in the DIS experiments.
The SU(3) symmetry relations \eqref{eIII} are no longer valid. However, the weaker 
relation obtained from current algebra \cite{HJL94}
still holds.
\beq 
{{G_A}\over{G_V}}(\Sigma^-\rightarrow n) =
{{\bra {n}  \Delta u - \Delta s \ket {n}  -
\bra {\Sigma^-}  \Delta u - \Delta s \ket {\Sigma^-}}
\over{2}}
\label{WW1.6}
\eeq
Relation (\ref{WW1.6}) is the SU(3) analogue of the familiar 
isospin relation
\beq 
{{G_A}\over{G_V}}(n\rightarrow p) =
{{\bra {p}  \Delta u - \Delta d \ket {p}  -
\bra {n}  \Delta u - \Delta d \ket {n}}
\over{2}}
\label{WW1a.6}
\eeq

 \section{Getting $\Delta u $, $\Delta d $  and $\Delta s $
 From Data }
 We have seen that two of the three parameters needed to determine the
three quantities $\Delta u $, $\Delta d $  and $\Delta s $ are obtainable 
from the experimental data on deep inelastic scattering of polarized leptons
on the proton and from the value of $G_A/G_V$ interpreted via the Bjorken sum
rule for the neutron decay.
There are several ways to continue. We first note that we can combine
\eqref{QQ1.1a}
and \eqref{QQ1.1b} to project out an isoscalar component of the spin structure
functions
\beq
\Delta u + \Delta d + (2/5)\cdot \Delta s = 0.333
\label{QQ1.1c}
\eeq
 
The conventional procedure for obtaining the needed additional experimental 
number to define three quantities is to use data from weak hyperon decays, 
interpreted using SU(3) symmetry via eqs. (3). 
by what is effectively an SU(3)-flavor rotation of the Bjorken sum rule.
This procedure has the advantage of dealing only with the three parameters
$\Delta u (p)$, $\Delta d (p) $  and $\Delta s (p)$ which are the total
contributions of quark spins of each flavor to the proton spin. There is no
need to break up these contributions into valence and sea contributions nor
to quark and antiquark.
 
However, flavor SU(3) is known to be
broken by quark mass differences which suppress the number of $s \bar s$ pairs
created by gluons in the sea relative to the number of $u \bar u$ and $d \bar d$
pairs. This has been borne out by neutrino experiments which suggest a
suppression factor of  roughly 2. 
 
We are thus led to breaking up the quark contributions into
valence and sea contributions. This is required on the one hand to provide a
mechanism for taking into account the SU(3) symmetry-breaking in the sea and
also to provide a description of the experiments which specifically measure the
antiquark content in the sea.
 
At this stage we wish to avoid a proliferation of models each with many
different ad hoc assumptions and many free parameters. We find that this can be
done in two ways (1) the conventional use of the hyperon weak decays; (2) a new
approach using the ratio of the proton and neutron magnetic moments. In both
cases we use the model discussed above in which the sea is not necessarily
flavor symmetric. 

Method (1) assumes, as in the discussion of the model, that the sea is a 
spectator in the weak transitions. Method (2) assumes 
that the sea is a spectator in the determination of the nucleon magnetic 
moments. 

Both assumptions can be questioned and justified only by hand-waving at this 
stage. The hand-waving for method (1) points to the success
of the model for Cabibbo theory and the observation that the contribution from 
a sea which violates flavor symmetry by a factor of two must be minimum. 
The hand-waving for method (2) notes that 
since quarks and antiquarks of the same flavor contribute with opposite
signs to magnetic moments, it is reasonable to assume a cancellation
between the integrals of quark and antiquark
momentum distributions which contribute to the magnetic moment.
This can be true even if there is a large flavor asymmetry in the sea
implied by the observed experimental violation of the Gottfried sum rule
\cite{Gottfried}.
 
What is particularly interesting is that each of the two approaches makes
assumptions that can be questioned, but that although these assumptions are
qualitatively very different, both give very similar results. 
The use of hyperon data
requires a symmetry assumption between nucleon and hyperon wave functions, which
is not needed for the magnetic moment method. But the use of magnetic moments
requires that the sea contribution to the magnetic moments be negligible, which
is not needed for the hyperon decay method.
 
We explore both approaches and two
possibilities for the strange quark content of the sea:\nl
(1) that the sea is SU(3) symmetric, 
\nl
(2) that the baryon wave function is described by 
eq.~\eqref{WW1.2} and the 
strange quark contribution differs from the
nonstrange  in the manner described by the parameter $\epsilon$ 
\beq
(1 + \epsilon) \Delta s_S= \Delta u_S =\Delta d_S
\label{QQ1.1d}
\eeq
proposed in the model \cite{PJECH},
which incorporates the weak decay data with the polarized 
DIS results and also maintains the good results of Cabibbo theory for weak 
decays.
 
\subsection*{A. The use of Hyperon Decay Data and SU(3) }
\subsubsection*{ 1. With a flavor-symmetric sea.}

 The standard analysis obtains an additional function of $\Delta u (p)$,
$\Delta d (p) $  and $\Delta s (p)$ from hyperon weak decay data. Rather than
using the SU(3) analysis with the D and F parametrization, we use a
mathematically equivalent formulation which is more transparent physically and
more easily extended to introduce SU(3) breaking.
The best fit to the isoscalar octet linear combination of $\Delta u (p)$,
$\Delta d (p) $  and $\Delta s (p)$ obtained from hyperon data gives.
\beq
\Delta u +  \Delta d - 2 \Delta s = 0.6
\label{QQ1.2}
\eeq
We now note that the relations obtained from the weak decay data
\eqref{QQ1.1b} and \eqref{QQ1.2}
depend only upon the valence quarks if the sea is SU(3)-symmetric.
Since there are no valence strange quarks in the proton, we obtain
\alpheqn
\beq
\Delta d_V = {-}0.33 \label{QQ1.3a}
\eeq
\beq
\Delta u_V = \phantom{{-}}0.93 \label{QQ1.3b}
\eeq
\beq
\Delta u_S + \Delta d_S + (2/5)\cdot \Delta s = -0.27
\label{QQ1.3c}
\eeq
\reseteqn
where the subscripts $V$ and $S$ denote valence and sea.
 
If we now  assume that the sea is SU(3) flavor-symmetric ($\epsilon = 0$) and
substitute into eqs.~\eqref{QQ1.1d} and \eqref{QQ1.3c},
 we immediately obtain
\alpheqn
\beq
\Delta u_S = \Delta d_S = \Delta s_S  = \Delta s =-0.11
\label{QQ1.4a}
\eeq
Thus
\beq
\Delta d = -0.44 \label{QQ1.4b}
\eeq
\beq
\Delta u = \phantom{{-}}0.82 \label{QQ1.4c}
\eeq
\beq
\Delta \Sigma = \phantom{{-}}0.27 \label{QQ1.4d}
\eeq
\reseteqn
These are the conventional values obtained by the mathematically equivalent D
and F parametrization.
 
\subsubsection*{2. With SU(3) for Valence Quarks and Breaking in the Sea}
 
We now assume that only the valence quarks contribute to weak decays and that
the sea is not SU(3) symmetric but is still the same for all octet baryons,
then
eqs.~\eqref{QQ1.1c}, \eqref{QQ1.3a} and \eqref{QQ1.3b} are still valid.
If we now assume
that $\Delta s$ in the sea is suppressed by a factor of 2 relative to
$\Delta u$ and $\Delta d$ ($\epsilon = 1$) we obtain instead of
\eqref{QQ1.4a},
\beq
\Delta u_S = \Delta d_S = 2\cdot \Delta s_S = -0.12 \label{QQ1.5}
\eeq
Thus
\alpheqn
\beq
\Delta d = -0.45 \label{QQ1.6a}
\eeq
\beq
\Delta u = \phantom{{-}}0.81
\label{QQ1.6b}
\eeq
\beq
\Delta \Sigma = \phantom{{-}}0.30
\label{QQ1.6c}
\eeq
\reseteqn
 
This is the well known result that the values of $ \Delta u$ and  $\Delta d$
obtained from the standard analysis of the data are insensitive to SU(3) breaking
in the sea, and only $ \Delta s$ is changed.
 
\subsection*{B. The use of Nucleon Magnetic Moment Data}
 
\subsubsection*{1. With a flavor-symmetric sea}
 
Rather than using hyperon weak decay data and assuming SU(3) symmetry,
we can obtain the needed alternative experimental input from nucleon
magnetic moments, under the assumption that these are proportional to
the valence quark contributions to the nucleon spin, 
$ \Delta d_V $ and $ \Delta u_V $, 
weighted by the quark charges, and using isospin symmetry
to relate proton and neutron wave functions.\footnote{That the values 
of $ \Delta d_V $ and $ \Delta u_V $ obtained from several
models fit the SU(6) prediction -(3/2) for the magnetic moment ratio to
better than 10\% under this assumption has been noted in \protect\cite{Qiu}.}

The neglect of sea contributions might be justified because their
quark and antiquark contributions to the magnetic moments have opposite sign
and tend to cancel\footnote{Clearly, the total numbers of quarks and
antiquarks of a given flavor are equal, but in order
for the cancellation to occur, the integrals over $x$
of the corresponding polarization distributions must be equal as well.}
We then obtain
\alpheqn
\beq
{{\mu_p}\over{\mu_n}} =
{{2 \Delta u_V (p) - \Delta d_V (p)}\over{2 \Delta u_V (n) - \Delta d_V (n)}} =
{{2 \Delta u_V (p) - \Delta d_V (p)}\over{2 \Delta d_V (p) - \Delta u_V (p)}} =
- {{2.79}\over{1.91}}
 \label{QQ1.7a}
\eeq
This gives
\beq
{{\Delta u_V (p) }\over{\Delta d_V (p) }} = - 3.56
\label{QQ1.7b}
\eeq
Assuming that polarizations of the light sea quarks are equal,
 $\Delta u_S = \Delta d_S$, their respective
contributions in eq.~\eqref{QQ1.1b} cancel each other,
and we can now solve eqs.~\eqref{QQ1.1b} and \eqref{QQ1.7b}
for $ \Delta d_V$ and $ \Delta u_V$, obtaining
\beq
\Delta d_V = -0.28 \label{QQ1.7c}
\eeq
\beq
\Delta u_V = \phantom{{-}}0.98
\label{QQ1.7d}
\eeq
\beq
\Delta u_S + \Delta d_S + (2/5)\cdot \Delta s =- 0.37
\label{QQ1.7e}
\eeq
\reseteqn
If we now  assume that the sea is SU(3) flavor-symmetric
($\epsilon = 0$) we
immediately obtain
\alpheqn
\beq
\Delta u_S = \Delta d_S = \Delta s_S = -0.16
\label{QQ1.8}
\eeq
\beq
\Delta d = -0.43
\label{QQ1.8b}
\eeq
\beq
\Delta u = \phantom{{-}}0.83 \label{QQ1.8c}
\eeq
\beq
\Delta \Sigma = \phantom{{-}}0.24 \label{QQ1.8d}
\eeq
\reseteqn
 
\subsubsection*{2. With flavor-symmetric breaking in the sea}
 
 If we now assume that $\Delta s$ in the sea is suppressed by a factor of 2
relative to $\Delta u$ and $\Delta d$ ($\epsilon = 1$) we obtain instead of
\eqref{QQ1.8}
\alpheqn
\beq
\Delta u_S = \Delta d_S = 2\cdot \Delta s_S = -0.17
\label{QQ1.9a}
\eeq
\beq
\Delta d = -0.45  \label{QQ1.9b}
\eeq
\beq
\Delta u = \phantom{{-}}0.81 \label{QQ1.9c}
\eeq
\beq
\Delta \Sigma = \phantom{{-}}0.28 \label{QQ1.9d}
\eeq
 
We thus see that the results for $\Delta u$ and $\Delta d$ remain essentially
the same, independently of whether the additional data are obtained from hyperon
decays or magnetic moments, and of whether the sea is flavor symmetric or the
strange quark contribution is reduced by a factor of two. Only $\Delta s$ is
changed.
 
\section*{Conclusions}
We now see that the results for the contributions of the nonstrange
quarks, $\Delta u$ and $\Delta d$, are determined primarily by the DIS
scattering and by the neutron decay and are essentially independent
of whether hyperon weak decay or nucleon magnetic moment data are used
to provide a third experimental input, and
whether one assumes an exact or seriously broken flavor SU(3) symmetry.

The strange
quark contribution $\Delta s$ is ${-}0.11$ when hyperon decay is used and 
${-}0.16$
when magnetic moments are used with an SU(3) symmetric sea. Both are
reduced by a factor of roughly two when the strange sea 
is reduced by a factor of two relative to the nonstrange sea
($\epsilon = 1$). But in any case all results are consistent within
two standard deviations of the value $\Delta s = -0.1$ if we assume an
experimental error of 25\%. Since the SU(3)-breaking factor of two is
determined only from measurement of unpolarized structure functions,
it is of interest to find other experiments which measure $\Delta s$
directly with greater precision.

The valence quark contributions to  $\Delta u_V$ and $\Delta d_V$ differ by
0.05, depending upon whether hyperon data or nucleon magnetic moments are
used to determine them. This  can clearly be attributed to the difference
in validity of the underlying assumptions in the two cases. Nevertheless,
the
difference between the values of the total contributions of $\Delta
u$ and $\Delta d$ to the proton spin is much smaller, 0.01. 
The values of $\Delta\Sigma$ obtained in the two methods differ by
0.03 or less.
What is
remarkable here is that these differences are so small considering that
their underlying assumptions are so different.
This effect is illustrated in Figure~1, where we plot $\Delta\Sigma$
and $\Delta s$ extracted in the two approaches, for somewhat wider range
of the strangeness suppression parameter, $0 \le \epsilon \le 3$. 

The question how flavor symmetry is broken remains open. We have pointed
out that model builders must keep track of how proposed SU(3) symmetry
breaking effects may effect the good SU(3) results for hyperon decays
obtained from Cabibbo theory and confirmed by experiment.  The observed
violation of the Gottfried sum rule remains to be clarified, along with
the experimental question of whether this violation of $\bar u - \bar
d$ flavor symmetry in the nucleon exists for polarized as well as for
unpolarized structure functions. The question of how SU(3) symmetry is
broken in the baryon octet can be clarified by experimental measurements
of $\Lambda$ polarization in various ongoing experiments
\cite{Lambdas}.

\bigskip
\begin{flushleft}
{\Large \bf Acknowledgements}
\end{flushleft}
One of us (HJL) thanks Gordon P. Ramsey for interesting discussions.
This   research was supported in part
by a grant from the United States-Israel
Binational Science Foundation (BSF), Jerusalem, Israel,
and by the Basic Research Foundation administered by the
Israel Academy of Sciences and Humanities.
\bigskip

\begin{figure}
%
\centerline{\epsfig{file=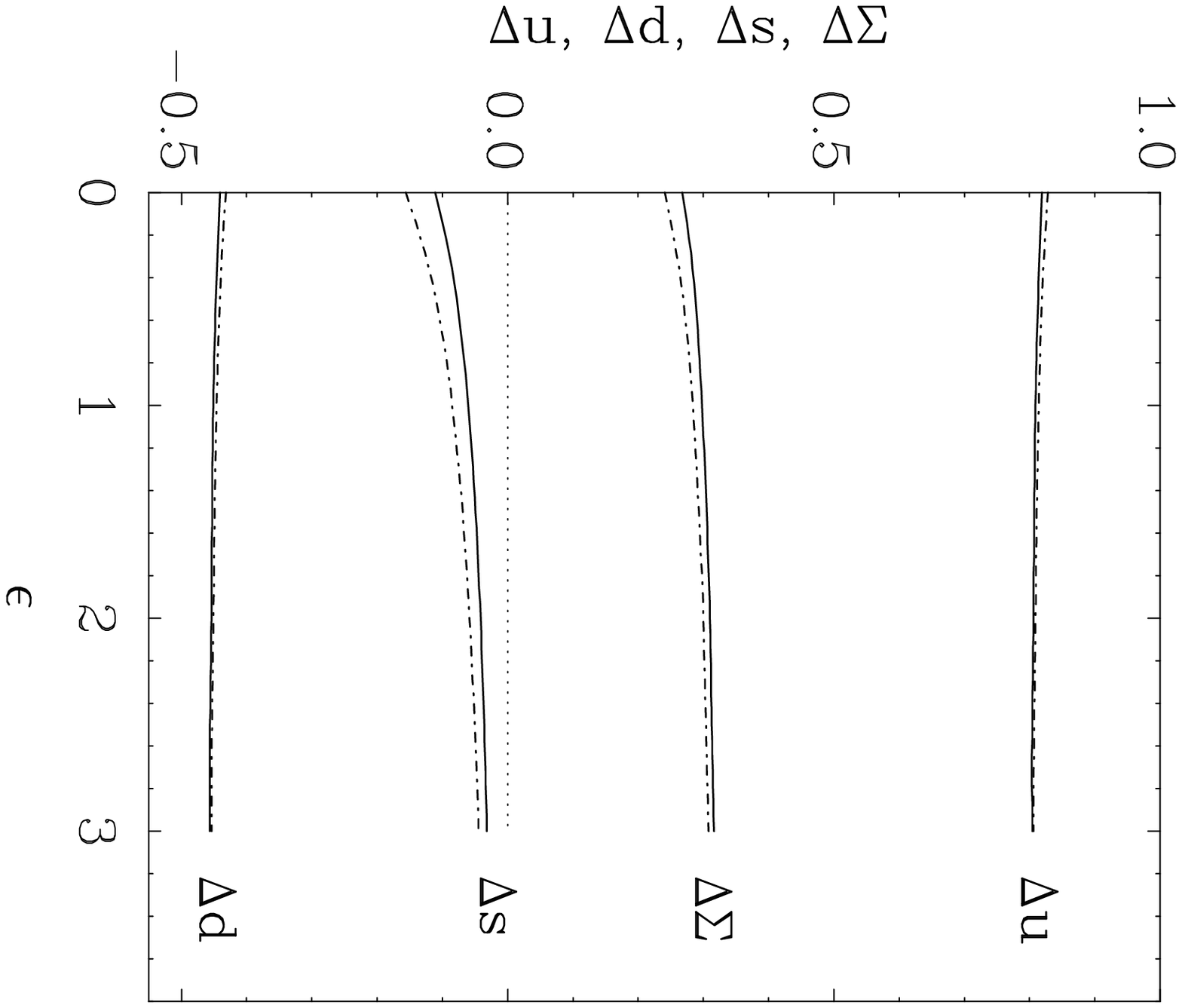,width=14cm,angle=90}}
\vspace*{1.5cm}
\hspace*{1cm}
\vbox{
{\bf Fig. 1.} 
$\Delta u$, $\Delta d$, $\Delta s$ and $\Delta \Sigma$ 
as function of $\epsilon$, using 
hyperon data (continuous line) and using ratio of magnetic moments
(dash-dotted line).
}
\end{figure}


\newpage
 
\begin{thebibliography}{99}
 
 
 
\bibitem{SMC93a} SMC, B. Adeva et al., Phys. Lett. B302 (1993) 533.
 
\bibitem{E142}  E-142, P.L. Anthony et al.,
                Phys. Rev. Lett. 71 (1993) 959.
\bibitem{SMC94b} SMC, D. Adams et al.,
                 Phys. Lett. B329 (1994) 399.
\bibitem{newpoldata}  
E143 Collab., K. Abe et al., Phys. Rev. Lett. 74 (1995) 346
and
Phys. Rev. {\bf D58} (1998)112003;
%
Spin Muon Collab., B. Adeva et al., 
Phys. Lett. {\bf B412}, 414 (1997),
%
%
Phys. Rev. {\bf D58} (1998)112002, {\em ibid.} {\bf D58} (1998)112001;
%
HERMES Collab. A. Airapetian et al., hep-ex/9807015,
Phys. Lett. {\bf B442} (1998) 484;
%
E154 Collab., K. Abe et al., hep-ph/9705344
Phys. Lett. {\bf B405} (1997) 180;
%
E155 Collab., P.L. Anthony et al., hep-ex/9904002
and references therein. 




 
\bibitem{As88} EMC, J. Ashman et  al.,
               Phys. Lett. B206, (1988) 364;
               Nucl. Phys. B328 (1989) 1.
 
\bibitem{Cl93}  F. E. Close and R. G. Roberts,
                 Phys. Lett. {\bf B 316} (1993) 165
 
\bibitem{Hs88}  S. Y. Hsueh {\it et al.},
                Phys. Rev.  D38, (1988) 2056.
		
\bibitem{Erice95} J. Ellis and M. Karliner,
Invited Lectures at the {\em Int. School of Nucleon Spin Structure}, 
Erice 1995, hep-ph/9601280, published in  Proceedings, B.~Frois, V.W.~Hughes,
N.~de Groot, Eds., World Scientific, 1997.

\bibitem{mk99} M. Karliner, unpublished (1999).

\bibitem{piNfluct}
A.W. Thomas, Phys. Lett. {\bf 126B} (1983) 97;
E.J. Eichten, I. Hinchliffe and C. Quigg,
Phys. Rev. {\bf D45} (1992) 2269;
E866 Collab., J. C. Peng et al., hep-ph/9804288,
Phys. Rev. {\bf D58} (1998) 092004;
for a recent review, see A.T.~Doyle, {\em Structure Functions},
hep-9812029,
plenary talk at XXIX ICHEP, Vancouver, July 1998.

\bibitem{Gottfried} 
NMC Collab., M. Arneodo et al., Phys. Rev. {\bf D50} (1994) 1;\nl
%
E866 Collab., E.A. Hawker et al., Phys. Rev. Lett. {\bf 80} (1998) 3715;\nl
%
Hermes Collab., K. Ackerstaff et al., hep-ex/9807013,
Phys. Rev. Lett. {\bf 81} (1998) 5519.

\bibitem{PJECH}Jechiel Lichtenstadt and Harry J. Lipkin,
Physics Letters B353 (1995) 119
 
\bibitem{CCFR} 
CCFR Collaboration,
A.O.~Bazarko {\it et al.},
hep-ex/9406007, Z. Phys. {\bf C65}, 189 (1995); 
see also the talk by Doyle in Ref.~\protect\cite{piNfluct}.
 
\bibitem{HJL94}  Harry J. Lipkin,
                 Phys. Lett.  B 337 (1994) 157.

\bibitem{Qiu} Jianwei Qiu, Gordon P. Ramsey, David Richards and Dennis
Sivers, Phys. Rev.  D41, (1990) 65.

\bibitem{Lambdas}
Fermilab E665 Collaboration, D.~Ashery, private
communication;\nl
OPAL Collaboration (K. Ackerstaff et al.), Eur. Phys. J. C2 (1998) 49,
 hep-ex/9708027;\nl
ALEPH Collaboration (D. Buskulic et al.), Phys. Lett. B374 (1996) 319;\nl
D.~de Florian (CERN), M.~Stratmann and W.~Vogelsang,
{\em Polarized $\Lambda$ production at HERA}, hep-ph/9710410;\nl
HERMES Collab. A. Airapetian et al., in \protect\cite{newpoldata}.



 
 
%
\end{thebibliography}
\end {document}